\documentstyle[12pt,A4]{article}

\input epsf
\epsfverbosetrue
\begin{document}
\begin{titlepage}
\begin{flushright}
HU-TFT-96-33\\
\today
\end{flushright}

\begin{centering}
\vfill
{\Large\bf Continuum extrapolation of energies of a four-quark system in 
lattice gauge theory}
\vspace{1cm}

Petrus Pennanen\footnote[1]{E-mail: {\tt Petrus.Pennanen@helsinki.fi}} 
\vspace{0.25cm}

{\em Research Institute for Theoretical Physics \\
P.O. Box 9 \\ FIN-00014 University of Helsinki \\ Finland }
\vspace{1.5cm}

{\bf Abstract}

\vspace{0.5cm}
\end{centering}

\noindent
A continuum extrapolation of static two- and four-quark energies calculated in 
quenched SU(2) lattice Monte Carlo is carried out based on Sommer's method of 
setting the scale. The $\beta$-function is obtained as a side product of the 
extrapolations. Four-quark binding energies are found to be essentially
constant at $\beta \ge 2.35$ unlike the two-body potentials. A model for 
four-quark energies, with explicit gluonic degrees of freedom removed, is 
fitted to these energies and the behaviour of the parameters of the model is 
investigated. An extension of the model using the first excited states of the 
two-body gluon field as additional basis states is found to be necessary for 
quarks at the corners of regular tetrahedra. 

\vfill

\end{titlepage}

\section{Introduction}

Systems of many hadrons play a crucial role in nature and it is important
to understand the hadronic interactions from first principles. This would shed 
light on multiquark bound states and e.g. meson-meson scattering. A convenient 
tool for analysing the problem would be a potential model for multi-quark 
systems with
explicit gluonic degrees of freedom removed. Such a model might be based on an 
effective two-body interaction as in the case of valence electrons in metals 
and nucleons in nuclei.

\begin{figure}[htb]
\newcommand{\thlen}{\setlength{\unitlength}{0.75pt}}
\newcommand{\quarksTR}{\multiput(0,0)(40,60){2}{\circle*{6}}
        \multiput(60,-20)(40,60){2}{\circle*{6}} }
\newcommand{\quarksNP}{\multiput(0,0)(0,40){2}{\circle*{6}}
        \multiput(60,-20)(40,20){2}{\circle*{6}} }
\newcommand{\quarksT}{\multiput(0,0)(40,60){2}{\circle*{6}}
        \multiput(60,20)(40,-20){2}{\circle*{6}} }
\newcommand{\quarksR}{\multiput(0,0)(0,40){2}{\circle*{5}}
        \multiput(60,0)(0,40){2}{\circle*{5}} }
\newcommand{\cubelabels}[3]{\put(28,-12){\makebox(0,0)[tr]{#1}}
        \put(85,-12){\makebox(0,0)[tl]{#2}}
        \put(105,20){\makebox(0,0)[cl]{#3}} }
\thlen
\newsavebox{\axes}
\savebox{\axes}{ {\thinlines
        \put(0,0){\vector(0,1){65}}
        \put(0,0){\vector(2,1){70}}
        \put(0,0){\vector(3,-1){85}} }}
\newsavebox{\grid}
\savebox{\grid}{ {\thinlines
        \multiput(0,0)(40,20){2}{\line(0,1){40}}
        \multiput(60,-20)(40,20){2}{\line(0,1){40}}
        \multiput(0,0)(60,-20){2}{\line(2,1){40}}
        \multiput(0,40)(60,-20){2}{\line(2,1){40}}
        \multiput(0,0)(0,40){2}{\line(3,-1){60}}
        \multiput(40,20)(0,40){2}{\line(3,-1){60}} }}
\setlength{\unitlength}{0.95pt}
\begin{center}
\setlength{\tabcolsep}{13pt}
\renewcommand{\arraystretch}{3.5}
\begin{tabular}{cccc}
 & \fbox{A} & \fbox{B} & \fbox{C} \\
\raisebox{15pt}{(R)} &
\begin{picture}(60,60)
\quarksR
\put(30,48){\makebox(0,0)[b]{$x$}}
\put(22,50){\vector(-1,0){22}}
\put(37,50){\vector(1,0){23}}
\put(67,20){\makebox(0,0)[l]{$y$}}
\put(71,27){\vector(0,1){13}}
\put(71,13){\vector(0,-1){13}}
\put(8,-2){\makebox(0,0)[b]{3}}
\put(52,-2){\makebox(0,0)[b]{2}}
\put(8,35){\makebox(0,0)[b]{1}}
\put(52,35){\makebox(0,0)[b]{4}}
\thicklines
\multiput(0,0)(60,0){2}{\line(0,1){40}}
\end{picture} &
\begin{picture}(60,60)
\quarksR
\put(0,5){\makebox(0,0)[b]{3}}
\put(60,5){\makebox(0,0)[b]{2}}
\put(0,27){\makebox(0,0)[b]{1}}
\put(60,27){\makebox(0,0)[b]{4}}
\thicklines
\multiput(0,0)(0,40){2}{\line(1,0){60}}
\end{picture} &
\begin{picture}(60,60)
\quarksR
\put(0,5){\makebox(0,0)[b]{3}}
\put(60,5){\makebox(0,0)[b]{2}}
\put(0,27){\makebox(0,0)[b]{1}}
\put(60,27){\makebox(0,0)[b]{4}}
\thicklines
\put(0,0){\line(3,2){60}}
\put(0,40){\line(3,-2){60}}
\end{picture} \\

\thlen
\raisebox{40pt}{(TR)} &

\thlen
\begin{picture}(100,95)(0,-30)
\put(0,0){\usebox{\axes}}
\put(0,0){\usebox{\grid}}
\quarksTR
\cubelabels{$x$}{$y$}{$z$}
\thicklines
\put(0,0){\line(3,-1){60}}
\put(40,60){\line(3,-1){60}}
\end{picture} &

\thlen
\begin{picture}(100,95)(0,-30)
\put(0,0){\usebox{\axes}}
\put(0,0){\usebox{\grid}}
\quarksTR
\thicklines
\put(0,0){\line(2,3){40}}
\put(60,-20){\line(2,3){40}}
\end{picture} &

\thlen
\begin{picture}(100,95)(0,-30)
\put(0,0){\usebox{\axes}}
\put(0,0){\usebox{\grid}}
\quarksTR
\thicklines
\put(0,0){\line(5,2){100}}
\put(40,60){\line(1,-4){20}}
\end{picture} \\

\raisebox{15pt}{(L)} &
\begin{picture}(80,40)(0,-20)
\multiput(0,0)(30,0){2}{\circle*{5}}
\multiput(50,0)(30,0){2}{\circle*{5}}
\put(15,-4){\makebox(0,0)[t]{$d$}}
\put(65,-4){\makebox(0,0)[t]{$d$}}
\put(25,10){\makebox(0,0)[b]{$r$}}
\put(17,12){\vector(-1,0){17}}
\put(32,12){\vector(1,0){18}}
\thicklines
\multiput(0,0)(50,0){2}{\line(1,0){30}}
\end{picture} &
\begin{picture}(80,40)(0,-20)
\multiput(0,0)(80,0){2}{\circle*{5}}
\multiput(30,-10)(20,0){2}{\circle*{5}}
\thicklines
\put(0,0){\line(1,0){80}}
\put(30,-10){\line(1,0){20}}
\end{picture} &
\begin{picture}(80,40)(0,-20)
\multiput(0,0)(50,0){2}{\circle*{5}}
\multiput(30,-10)(50,0){2}{\circle*{5}}
\thicklines
\multiput(0,0)(30,-10){2}{\line(1,0){50}}
\end{picture} \\

\thlen
\raisebox{60pt}{(T)} &
\thlen
\begin{picture}(100,110)(0,-55)
\put(0,0){\usebox{\axes}}
\put(0,0){\usebox{\grid}}
\quarksT
\cubelabels{$r$}{$d$}{$d$}
\thicklines
\put(0,0){\line(3,1){60}}
\put(40,60){\line(1,-1){60}}
\end{picture} &

\thlen
\begin{picture}(100,110)(0,-55)
\put(0,0){\usebox{\axes}}
\put(0,0){\usebox{\grid}}
\quarksT
\thicklines
\put(0,0){\line(1,0){100}}
\put(60,20){\line(-1,2){20}}
\end{picture} &

\thlen
\begin{picture}(100,110)(0,-55)
\put(0,0){\usebox{\axes}}
\put(0,0){\usebox{\grid}}
\quarksT
\thicklines
\put(0,0){\line(2,3){40}}
\put(100,0){\line(-2,1){40}}
\end{picture}
\end{tabular}
\end{center}
\caption{Some of the simulated four-quark geometries and their two-body 
pairings. \label{fconf}}
\end{figure}
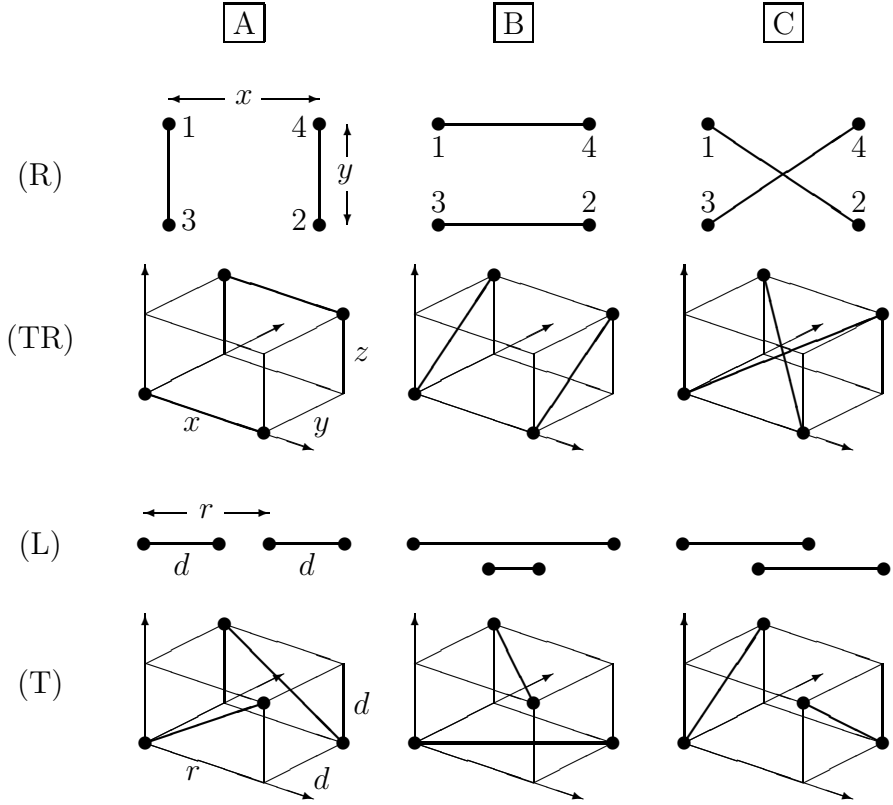

Perturbation theory of QCD cannot even treat the confinement of quarks and 
gluons into hadrons, and the only known way to make realistic calculations of 
interacting quark clusters are Monte Carlo lattice techniques. Green and 
coworkers have been simulating systems of four static quarks in quenched 
SU(2) mainly on $16^3\times 32$ lattices \cite[references therein]{glpm:96}.
Four quarks, because it is the smallest number that can be partitioned into 
different colour-singlet groups. Energies of several configurations such as 
rectangular (R), linear (L) and tetrahedral (T) -- examples of which are 
shown in figure \ref{fconf} --
have been simulated to get a set of geometries representative of the general 
case. The $\beta$ values used have been 2.4 and 2.5. 

However, e.g. Booth et al. \cite{boo:92} conclude that asymptotic scaling for 
SU(2) gauge theory begins at $\beta > 2.85$, so there is reason to suspect 
that the simulation results could contain significant lattice artefacts. 
Previous 
estimates show that finite size effects of our simulations are unimportant, 
while the conclusion that scaling (as opposed to asymptotic scaling) has been 
achieved at $\beta=2.4$ is somewhat more questionable for larger configurations
and excited states \cite{gmps:93}. It is therefore expected that a continuum 
extrapolation could possibly yield different 
energies and point out artefacts in a parameterization of the energies while 
making the physical content clearer. Such an extrapolation is the object 
of this work. 

The geometries to be simulated here were chosen to be squares and tilted 
rectangles 
[(R) with $x=y$ and (TR) in figure \ref{fconf}], because the simple model 
described below works for them and these geometries also exhibit the largest 
binding energies. The $\beta$ values (and lattice sizes) used were 
$\beta=2.35$, $2.4$ ($16^3\times 32$), 2.45 ($20^3\times 32$), 2.5 
($24^3\times 32$) and 2.55 ($26^3 \times 32$). In addition to these 
two-body runs were performed at $\beta=2.3$ ($12^3\times 32$) to set the scale
at a lower value of $\beta$.

\section{A model for four-quark energies}

A model for the energy of four static quarks with explicit gluonic degrees of 
freedom removed has been developed by Green and coworkers 
\cite[references therein]{glpm:96}. 
In this model a potential matrix is 
diagonalized with different two-body pairings as basis states. The basis 
states for some simulated configurations are shown in figure \ref{fconf}. 
For example, in the case of two basis states A and B, the eigenvalues 
$\lambda_i$
are obtained from
\begin{equation}
\label{Ham}
\left[{\bf V}-\lambda_i {\bf N}\right]\Psi_i=0, \label{emodel}
\end{equation}
with
\begin{equation}
\label{NV}
{\bf N}=\left(\begin{array}{ll}
1&f/N_c \\
f/N_c&1\end{array}\right)\ \ {\rm and}\ \ {\bf V}=\left(\begin{array}{cc}
v_{13}+v_{24} & \frac{f}{N_c}V_{AB}\\
\frac{f}{N_c}V_{BA}&v_{14}+v_{23}\end{array}\right), 
\end{equation}
where $v_{ij}$ represents the static two-body potential between quarks $i$ and 
$j$. $V_{AB}$ comes from the perturbative expression
\begin{equation}
V_{ij}=-{\cal N}(N_c) {\bf T_i} \cdot {\bf T_j} v_{ij}, \label{evij}
\end{equation}
where for a colour singlet state $[ij]^0$ the normalization is chosen to give\\
$<[ij]^0|V_{ij}|[ij]^0>=v_{ij}$. The four-quark 
binding energies $E_i$ are obtained by subtracting the 
internal energy of the basis state with the lowest energy, e.g. 
$$
E_i = \lambda_i-(v_{13}+v_{24}).
$$

A central 
element in the model is a phenomenological factor $f$ appearing in the 
overlap of the basis states $<A|B> = f/N_c$ for $SU(N_c)$. This factor is a 
function of  the spatial coordinates of all four quarks, making the 
off-diagonal 
elements of ${\bf V}$ in eq. \ref{NV} {\em four-body} potentials, and 
attempts to take into account the decrease of overlap from the weak coupling 
limit, where $<A|B>=1/N_c$. 
Perturbation theory to $O(\alpha^2)$ also produces the two-state model of eq. 
\ref{emodel} with $f=1$ \cite{lan:95}. 
Several parameterizations for $f$ have been suggested, a general form being
\begin{equation}
f = f_c e^{-k_A b_S A-k_P \sqrt{b_S} P}. \label{ef}
\end{equation}
Here $b_S$ is the string tension, $f_c$ a normalization constant and 
$k_A,k_P$ multiply the minimal area and perimeter bounded by the four quarks 
respectively. In this work either the normalization or the perimeter term
is omitted from eq. \ref{ef}. 

This simple version of the ``$f$-model''  works for quarks in the corners of 
squares and 
(tilted) rectangles [(R) and (TR) in fig. \ref{fconf}], but fails to predict 
some features of non-planar geometries, e.g. 
the doubly degenerate ground state energy of a regular tetrahedron 
[(T) with $r=d$]. 
In section \ref{sdev} there will be introduced a generalization of the model 
that is capable of reproducing this degeneracy.

\section{Extrapolating to the Continuum}

\subsection{Setting the scale}

Previously the scale in our simulations has been set by equating
to the experimental continuum value the string
tension $b_S = \lim_{r\rightarrow\infty} F(r)$, where $F(r)$ is the force 
between two static quarks, in lattice units. The string tension, however, 
was obtained by fitting the 
lattice parameterization of two-body potential to values from simulations at 
$r/a=2,\ldots,6$ and not as $r\rightarrow\infty$.

Sommer \cite{som:94} has designed a popular new way to set the scale that 
uses only intermediate distances. First
the force $F(r/a)$ between two static quarks at separation $r/a$ is
calculated. By solving
\begin{equation}
(r_0/a)^2 F(r_0/a) = c \label{esommer}
\end{equation}
with $c=1.65$ for $r_0/a$, we get the equivalent of the continuum value 
$r_0\approx 0.5$ fm in lattice units. The constant on the right-hand side of 
eq. 
\ref{esommer} has been chosen to correspond to a distance scale where we have 
best information on the force between static quarks. According to the widely 
cited paper of Buchm\"uller and Tye \cite{buc:81} various 
nonrelativistic effective potentials which successfully model heavy quarkonia 
agree in the radial region
0.1 to 1.0 fm and predict r.m.s. radii from 0.2 to 1.5 fm for 
$\bar{c}c$ and $\bar{b}b$ systems. Even though these effective 
potentials are not the same as the QCD potential between static quarks, the 
distance scale where we have the best experimental evidence seems to be 
around 0.5 fm. A different estimate is given by Leeb et al. \cite{lee:90}, 
who fit modified Cornell and Martin potentials to meson masses and claim that 
the known mesons determine the potential model-independently only between 
$r=0.7$ fm and $r=1.8$ fm. 

In the Cornell \cite{eic:80} and Richardson \cite{ric:79} potential models 
the constant 1.65 in eq. \ref{esommer} corresponds to $r_0 = 0.49$ fm.
On the other hand, the Martin model \cite{mar:81}, with the strange quark 
counted as heavy, gives $r_0 = 0.44$ fm and the modified Cornell and Martin 
potentials of ref. \cite{lee:90} result in 0.56 fm and 0.44 fm respectively 
(the published parameter values are incorrect 
\footnote{Personal communication with H. Leeb.}). 

Choosing $c=2.44$ as the scale setting constant makes the Richardson, modified 
Cornell and modified Martin potentials agree on $r_0 = 0.66$ fm with the basic
Cornell and Martin models giving $r_0=0.64$ fm and 0.625 fm respectively. These
models are clearly in better agreement at $c = 2.44$ than at $c = 1.65$. 
However, it is not excluded that the 
agreement may be accidental due to uncertainties in the models.

To set the scale, a lattice parameterization of the two-quark potential
was first fitted to values obtained from simulations at $r/a=2,\dots,6$ for 
each $\beta$. The parameterization used was
\begin{equation}
\label{coul1}
v_L(r)=-\left( \frac{e}{r} \right)_L + b_S r + v_0,
\end{equation}
where the on-axis lattice Coulomb potential is \cite{lan:82}
\begin{eqnarray}
\label{coul}
\left(\frac{1}{r}\right)_L & = & \pi a^3 \int_{-\pi/a}^{\pi/a}
\frac{dk_1dk_2dk_3}{(2\pi)^3}\frac{e^{-irk_1}}{\sum_{i=1}^{3}\sin^2(k_i a/2)}
\nonumber \\
& \approx & \frac{\pi}{L'^3}\sum_{\bar{q}}
\frac{\cos(rq_1)}{\sum_{i=1}^{3}\sin^2(q_i a/2)} \\
aq_i & = & -\pi,\frac{\pi}{L'},\ldots,\frac{\pi(2L'-1)}{L'}. 
\end{eqnarray}
Here $L'$ is twice the number of spatial sites along one axis. 
The fit 
results are presented in table \ref{t2bfits} for data from tilted rectangle 
and square geometry runs performed in Helsinki, and $\beta=2.5$ square runs 
from Wuppertal. 

\begin{table}[htb]
\begin{center}
\begin{tabular}{|l|c|c|c|c|} \hline
$\beta$  & $e$ & $b_Sa^2$ & $v_0a$ & $\chi^2$ \\ \hline
2.3      & 0.299(14) & 0.1278(20) & 0.588(11) & 2.8 \\ 
2.35     & 0.255(12) & 0.0985(15) & 0.557(9)  & 2.6 \\
2.35(TR) & 0.260(5)  & 0.0978(7)  & 0.562(4)  & 1.0 \\
2.4      & 0.248(7)  & 0.0709(10) & 0.554(6)  & 0.1 \\
2.4(TR)  & 0.238(9)  & 0.0716(13) & 0.546(7)  & 0.18 \\
2.45     & 0.244(5)  & 0.0494(7)  & 0.551(4)  & 0.04 \\
2.45(TR) & 0.238(5)  & 0.0507(6)  & 0.545(4)  & 0.05 \\
2.5      & 0.226(5)  & 0.0373(7)  & 0.531(4)  & 0.55 \\ 
2.5(TR)  & 0.233(4)  & 0.0367(5)  & 0.534(3)  & 0.53 \\
2.5[W]   & 0.234(4)  & 0.0371(4)  & 0.536(3)  & 0.11 \\
2.55     & 0.223(2)  & 0.0271(3)  & 0.522(2)  & 0.03 \\
2.55(TR) & 0.224(5)  & 0.0268(6)  & 0.523(4)  & 0.13 \\ \hline
\end{tabular}
\caption{Parameters of the lattice two-quark potential. [W] refers to data
from Wuppertal, (TR) to tilted rectangles. \label {t2bfits}}
\end{center}
\end{table}

The force was 
calculated from this parameterization by the finite difference
\begin{equation}
F(r_I) = \frac{V(r)-V(r-d)}{d}, \label{eforce}
\end{equation}
using $d=a$. Here
\begin{equation}
r_I = \sqrt{\frac{d}{[G(r-d)-G(r)]}} \\,
\end{equation}
where
\begin{equation}
G(r) = \frac{1}{a}\left(\frac{1}{r}\right)_L
\end{equation}
and $r_I$ has been defined to remove
$O[(a/r)^2]$ lattice artefacts from the argument of $F$, and to make
$F(r_I)$ a tree-level improved observable \cite{som:94}. 

After getting the force at points $r_I/a$, the expression $(r_I/a)^2 F(r_I/a)$ 
was interpolated to get $r_0/a$ corresponding to eq. \ref{esommer}. 
Results for different $\beta$'s are 
presented in table \ref{tri}, with lattice spacing $a$ corresponding to 
$r_0 = 0.49$ fm and $a^{\rm II}$ to $r_0^{\rm II} = 0.66$ fm. 
For comparison, a value of $a^{b_S}$ obtained from equating 
the string tension to the somewhat arbitrary continuum value 
$\sqrt{b_S}=0.44$ GeV is presented. 
In all cases the $a^{\rm II}$ consistently agrees with $a^{b_S}$, whereas $a$ 
disagrees, $a^{b_S}/a$ being
$\approx 1.09$ at each $\beta$. Choosing $\sqrt{b_S}=0.478(4)$ GeV would move
$a^{b_S}$ to the same value as $a$. A similar disagreement with the ratio 
of spacings being $\approx 1.04$ 
was found by the 
SESAM collaboration working in SU(3) and using $c=1.65$ in eq. \ref{esommer} 
\cite{gla:96}. However, one should 
keep in mind that the quenched SU(2) string tension does not have to equal a 
phenomenological value. 

As using $c=2.44$ in the Sommer scheme produces better agreement with various 
continuum potential models and leads to agreement of 
$a^{\rm II}$ and $a^{b_S}$ at continuum string tension value $b_S = 0.44$ 
GeV, it seems that $c=1.65$ underestimates the lattice spacing. Using 
$c=2.44$, corresponding to $r_0\approx 0.66$ fm, is in our case apparently a 
better choice than $c=1.65$ and was chosen for this work, although the 
possibility that the difference is accidental cannot be excluded.

The $\beta=2.5[W]$ values are calculated 
from potential data by a Wuppertal group doing simulations similar to ours 
\cite{bal:94}. The scales given by our and their data using the same analysis 
method agree well at $\beta=2.5$ , but the scales obtained from the 
published $b_s$ values [0.0826(14) fm  vs. 0.0864(5) fm] using the same data 
do not agree, which hints of the 
sensitivity of any intermediate distance determination of the string tension 
to the details of the fitting procedure. The errors in table \ref{tri} 
were estimated by adding or subtracting the one-sigma errors of the 
parameters of the lattice two-body potential, so systematic errors from e.g. 
the choice of $r_0$ are not included. 

\begin{table}[htb]
\begin{center}
\begin{tabular}{|l|c|c|c|c|c|} \hline
$\beta$  & $r_0/a$ & $a$ [fm] & $r_0^{\rm II}/a$  & $a^{\rm II}$ [fm] & 
$a^{b_S}$ [fm] \\ \hline
2.3      & 3.25(4) & 0.1508(20) & 4.09(5) & 0.1613(18)  & 0.1603(13) \\ 
2.35     & 3.76(2) & 0.1302(6)  & 4.71(5) & 0.1401(14)  & 0.1408(11)  \\
2.35(TR) & 3.77(2) & 0.1299(7)  & 4.72(3) & 0.1397(6)   & 0.1402(5)  \\
2.4      & 4.45(5) & 0.1101(10) & 5.56(5) & 0.1186(10)  & 0.1194(9) \\
2.4(TR)  & 4.44(2) & 0.1103(5)  & 5.55(2) & 0.1190(4)   & 0.1200(11)  \\
2.45     & 5.34(4) & 0.0918(7)  & 6.62(5) & 0.0997(8)   & 0.0997(7)  \\
2.45(TR) & 5.27(5) & 0.0929(7)  & 6.59(5) & 0.1002(7)   & 0.1010(6)  \\
2.5      & 6.18(5)  & 0.0793(5) & 7.71(8) & 0.0856(9)   & 0.0866(9)  \\ 
2.5(TR)  & 6.21(5)  & 0.0789(5) & 7.75(6) & 0.0851(6)   & 0.0859(6)  \\ 
2.5[W]   & 6.18(3)  & 0.0793(3) & 7.71(5) & 0.0856(6)   & 0.0864(5) \\
2.55     & 7.26(5)  & 0.0675(5) & 9.04(6) & 0.0730(5)   & 0.0741(7)  \\
2.55(TR) & 7.29(10) & 0.0673(9) & 9.08(12) & 0.0727(10) & 0.0734(8)  \\ 
\hline
\end{tabular}
\caption{Values of $r_0/a$ and $a$ for each $\beta$. 
$r_0^{\rm II}/a$ and $a^{\rm II}$  are calculated using $c=2.44$ in 
eq.\protect\ref{esommer}.}
\label{tri}
\end{center}
\end{table}

The values of $r_0$ obtained agree with the relation $r_0 = \sqrt{(c-e)/b_S}$ 
using $c$ from eq. \ref{esommer} and $e,b_S$ from the fit of the two-body 
parameterization of eq. \ref{coul1}.

A criterion for scaling of the two-body potential is equality of the parameter 
$e$ for fits in the same physical distance range. In our case the fit range 
is shorter at higher $\beta$'s, and the values in table \ref{t2bfits} support
scaling at $\beta=2.5$ and $\beta=2.55$.

In an attempt to take into account non-perturbative effects we can fit the 
values of $a$ using the perturbative three-loop relation between $a$ and
$\beta$ but with two extra terms, 
\begin{equation}
a = \frac{1}{\Lambda_{LAT}}e^{-\frac{\beta}{4N_c\beta_0}}(\frac{2N_c\beta_0}
{\beta})^{-\frac{\beta_1}{2\beta_0^2}}\left\{ 1+
\frac{\beta_1^2-\beta_2^L\beta_0}{2\beta_0^3}2N_c/\beta+A_2/\beta^2 
+ A_3/\beta^3 + O(1/\beta^4)\right\}. \label{eabeta}
\end{equation}
Here, as usual, $\beta_0=\frac{11}{3}\frac{N_c}{16\pi^2}$ and 
$\beta_1=\frac{34}{3}(\frac{N_c}{16\pi^2})^2$, while the recently calculated
$\beta_2^L=(\frac{N_c}{16\pi^2})^3(-366.2+\frac{1433.8}{N_c^2}-
\frac{2143}{N_c^4})$ \cite{all:96}.  Using Michael's result from 
ref. \cite{mic:92}, let us fix
$\sqrt{b_S}/\Lambda_{LAT}=31.9$.
Fitting to all the values of $a^{II}$ in table \ref{tri} results in $A_2=-6.72(73), 
A_3=25.8(1.8)$ with $\chi^2=0.9$. The $\chi^2$ value is per d.o.f. as 
elsewhere in this paper. A modification of eq. \ref{eabeta} with no two to 
fourth order terms in $1/\beta$ but only a fifth order term with coefficient
$A_5=55.7(3)$ gives also a low $\chi^2=1.1$. 

The $\beta$-function $b \equiv \partial \beta/\partial \ln a$ obtained using
the former fit is presented in table \ref{tbeta} with comparisons to 
three-loop predictions and also estimates by Engels et al.
using a thermodynamic approach \cite{eng:95}. The errors in our estimates 
were obtained by fitting to the values of $a$ instead of $a^{II}$ in table 
\ref{tri} and using the form of eq. \ref{eabeta} with only fifth order 
coefficient $A_5$ fitted in order to account for systematic errors from the
choice of $c$ in eq. \ref{esommer} and the assumed functional dependence in 
eq. \ref{eabeta}.

The agreement with Engels et al. is good except at the lowest 
$\beta$ -- the decrease in $b$ with decreasing $\beta$ at $\beta \leq 2.37$
is not reproduced by our approach. An estimate of $b=-0.35(2)$ at $\beta=2.4$
obtained using energy sum-rules \cite{gre:96} seems somewhat low when
compared to the values in table \ref{tbeta}. 

At $\beta=2.5$ a fit of the two-loop analogue of eq. \ref{eabeta} gives
a value 78\% of the perturbative two-loop prediction, which is modestly
improved to 80\% for the corresponding three-loop expressions. At these 
values of the bare coupling, additional terms in the perturbative series are 
thus not likely to lead to a significantly better agreement with PT. 

\begin{table}[htb]
\begin{center}
\begin{tabular}{|l|c|c|c|c|} \hline
$\beta$  & $b$ from eq. \ref{eabeta} & ref. \cite{eng:95} & 3-loop PT \\ \hline
2.3      & -0.298(9)   & -0.3393  & -0.3898 \\ 
2.35     & -0.301(6)   & -0.3055  & -0.3896 \\
2.4      & -0.305(6)   & -0.3018  & -0.3893 \\
2.45     & -0.308(3)   & -0.3057  & -0.3891 \\
2.5      & -0.312(2)   & -0.3115  & -0.3889 \\ 
2.55     & -0.315(4)   & -0.3183  & -0.3887 \\ \hline
\end{tabular}
\caption{Comparison between values of $b \equiv \partial \beta/\partial \ln a$ from different methods. \label {tbeta}}
\end{center}
\end{table}

\subsection{Extrapolating two-body potentials \label{2exp}}

\begin{figure}[p]
\hspace{0cm}\epsfxsize=400pt\epsfbox{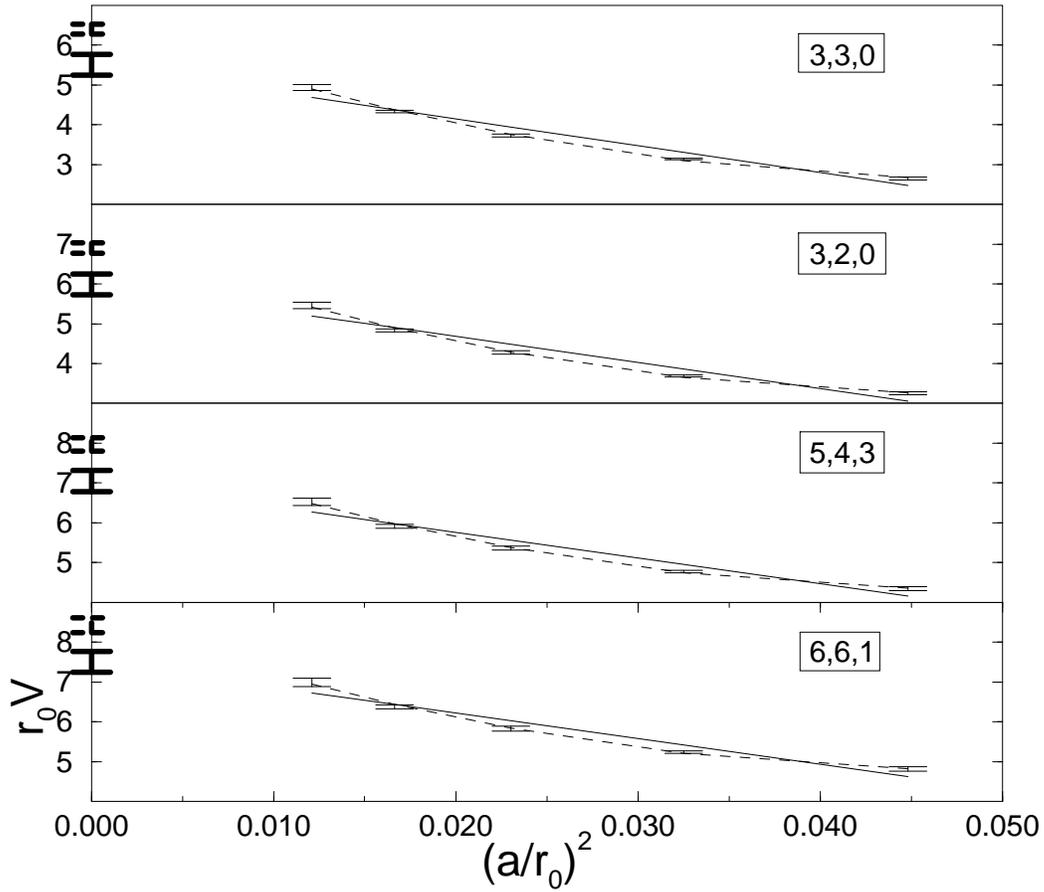}
\caption{$r_0V$ vs. $(a/r_0)^2$ for some two-body potentials, dimensions 
$x,y,z$ shown in lattice units at $\beta=2.5$}
\label{f2bext}
\end{figure}

To extrapolate to the continuum, we need values of energies 
at different $\beta$'s but corresponding to the same physical size. 
Two-body potentials were interpolated from the values given by the lattice 
parameterization. Figure \ref{f2bext} shows an extrapolation of some of the  
two-body potentials involved in tilted rectangles. Others (from a total of 29)
look similar -- more pictures can be seen in ref. \cite{pp:97}.

Linear (quadratic) fits with all data points included give $\chi^2$ values 
from 15 to 35 (1.8 to 3.5), while cutting the $\beta=2.35$ point off improves
these to from 8 to 17 (0.5 to 1.5). Continuum energies given by the latter
quadratic extrapolations are 6-8\% higher than by the quadratic extrapolations
including the $\beta=2.35$ data. 

The 29 quadratically extrapolated continuum potentials can be parameterized 
using eq. \ref{coul1} with the lattice coulomb term replaced by $e/r$ and
$e=0.21(11),\, b_S = 5.11(80)$ fm$^{-2}$ and $v_0 = 9.38(11)$ fm$^{-1}$ 
($\chi^2 = 0.4$). The value of the string tension corresponds to 
$\sqrt{b_S}=0.446(37)$ GeV, which agrees well with the continuum value used to 
obtain the $a^{b_S}$ in table \ref{tri}.

The lattice artefact at each $\beta$, i.e. difference from quadratically 
extrapolated continuum energy, has a linear dependence on the inverse 
separation of the quarks in lattice units squared $(a/R)^2$. The slopes of 
this dependence at each $\beta$ are within errors of each other, the average 
being 0.9(2).

\subsection{Extrapolating the binding energies}

The continuum extrapolations were performed for the square and tilted rectangle
geometries, as the simple $f$-model works for them and 
allows the interpolation of binding energies to distances which are non-integer
multiples of the lattice spacing. The ground state energy is given by
\begin{eqnarray}
E_0 & = &  \{2v_{s1} - 2v_{s2} + f^2(v_{s2}-v_d) \\
    &  & \mbox{} \frac{+\sqrt{[2v_{s2} - 2v_{s1} - f^2(v_{s2}-v_d)]^2 -
          f^2(-4 + f^2)(v_d - v_{s2})^2}\} }
     {-2 + 0.5f^2} \nonumber,
\end{eqnarray}
where $v_{s1},v_{s2},v_d$ denote the two-body potentials along the shorter and
longer side and the diagonal of the rectangle, respectively.

\begin{figure}[p]
\hspace{0cm}\epsfxsize=400pt\epsfbox{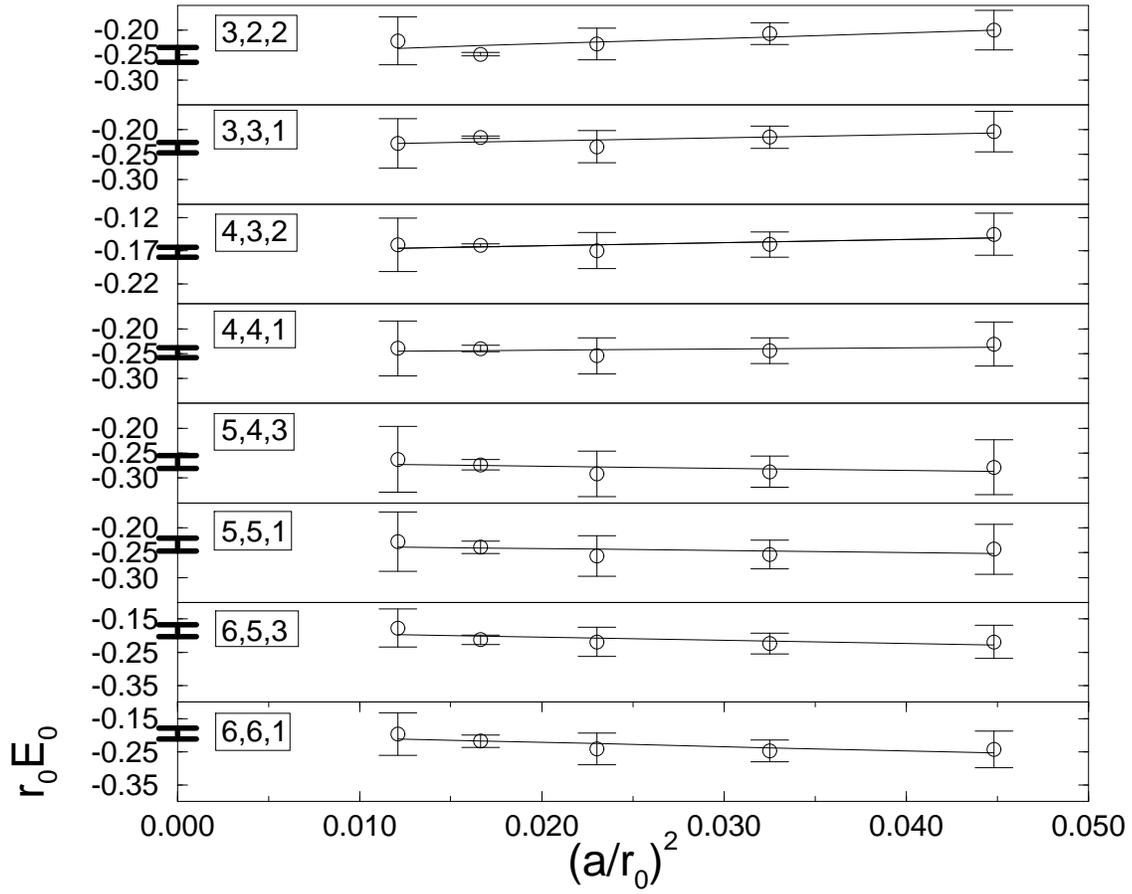}
\caption{$r_0E_0$ vs. $(a/r_0)^2$ for tilted rectangles with dimensions in 
lattice units shown at $\beta=2.5$}
\label{fsomext1}
\end{figure}

The two-parameter model of eq. \ref{ef} (with $f_c$, $k_A$ or $k_A$, $k_P$) 
was fitted to the energies of squares and tilted rectangles separately. 
Result of extrapolating from these values is presented in figure 
\ref{fsomext1} for tilted rectangles. Similar interpolations and 
extrapolations were performed for squares with sizes $2a,\ldots,6a$ at 
$\beta=2.5$. Slopes for 11 of the 13 extrapolated energies were consistent with
zero, supporting scaling of the binding energies at $\beta\geq 2.35$. 
Linear fits have low $\chi^2$ values and are more reliable than 
quadratic fits, which introduce an unnecessary extra parameter and lead to very
large errors on the continuum values.

\section{The $f$-model in the continuum}

\subsection{Extrapolation of the parameters of $f$}

The parameters of the factor $f$ were extrapolated from values given by a fit 
to the combined energies of squares and tilted rectangles, as shown in table 
\ref{t2parfc} for parameterizations with either the normalization or
perimeter term omitted from eq. \ref{ef}. 

\begin{table}[htb]
\begin{center}
\begin{tabular}{|l||c|c|c||c|c|c|c|} \hline
$\beta$       & $f_c$      & $k_A$      & $\chi^2$ & $k_A$ & $k_P$ & $\chi^2$ 
\\ \hline
2.55          & 0.974(17)  & 0.77(6)    & 0.9 & 0.71(10) & 0.020(16) & 1.0 \\
2.5           & 0.992(17)  & 0.76(4)    & 0.3 & 0.73(8) & 0.008(13) & 0.3 \\
2.45          & 0.925(10)  & 0.64(3)    & 1.0 & 0.49(4) & 0.057(8) & 1.2 \\
2.4           & 0.864(15)  & 0.57(2)    & 1.25 & 0.38(4) & 0.087(10) & 1.25 \\
2.35          & 0.843(8)   & 0.57(2)    & 1.4 & 0.36(3) & 0.101(5) & 1.1 \\ 
\hline
\end{tabular}
\caption{Two-parameter $f$-models fitted to the square and tilted rectangle 
data. \label{t2parfc}}
\end{center}
\end{table}

\begin{figure}[p]
\hspace{0cm}\epsfxsize=360pt\epsfbox{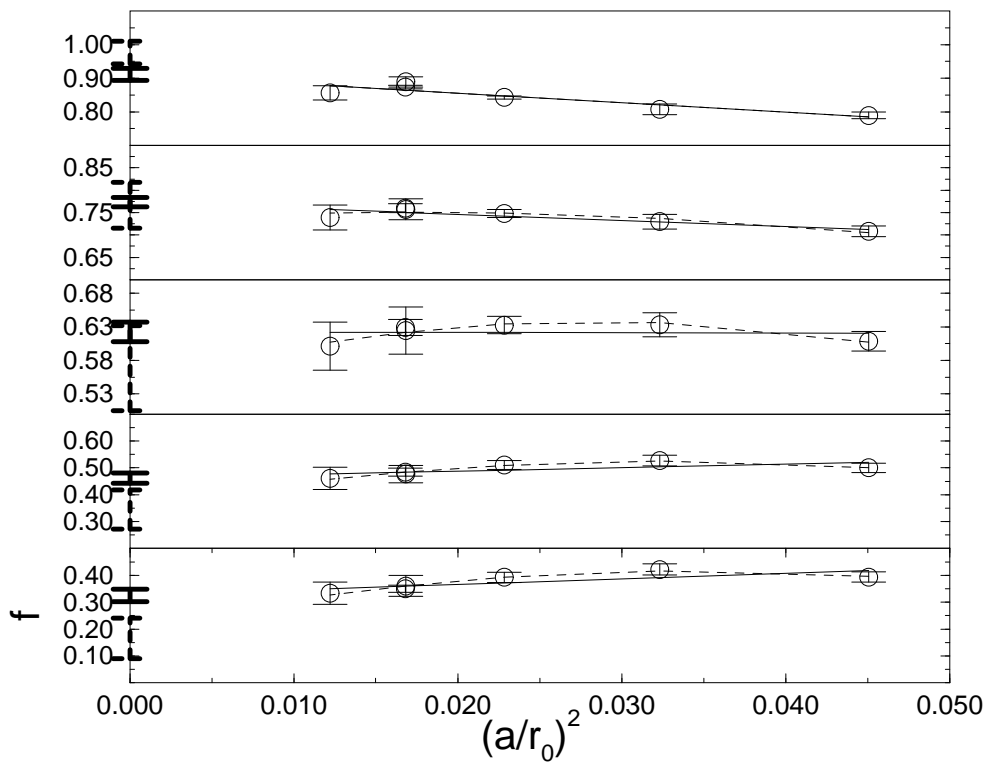}
\caption{A continuum extrapolation of $f$'s for squares with length of a side 
from $2a$ (top) to $6a$ (bottom) at $\beta=2.5$.}
\label{fsomf}
\end{figure}

\begin{figure}[p]
\hspace{0cm}\epsfxsize=360pt\epsfbox{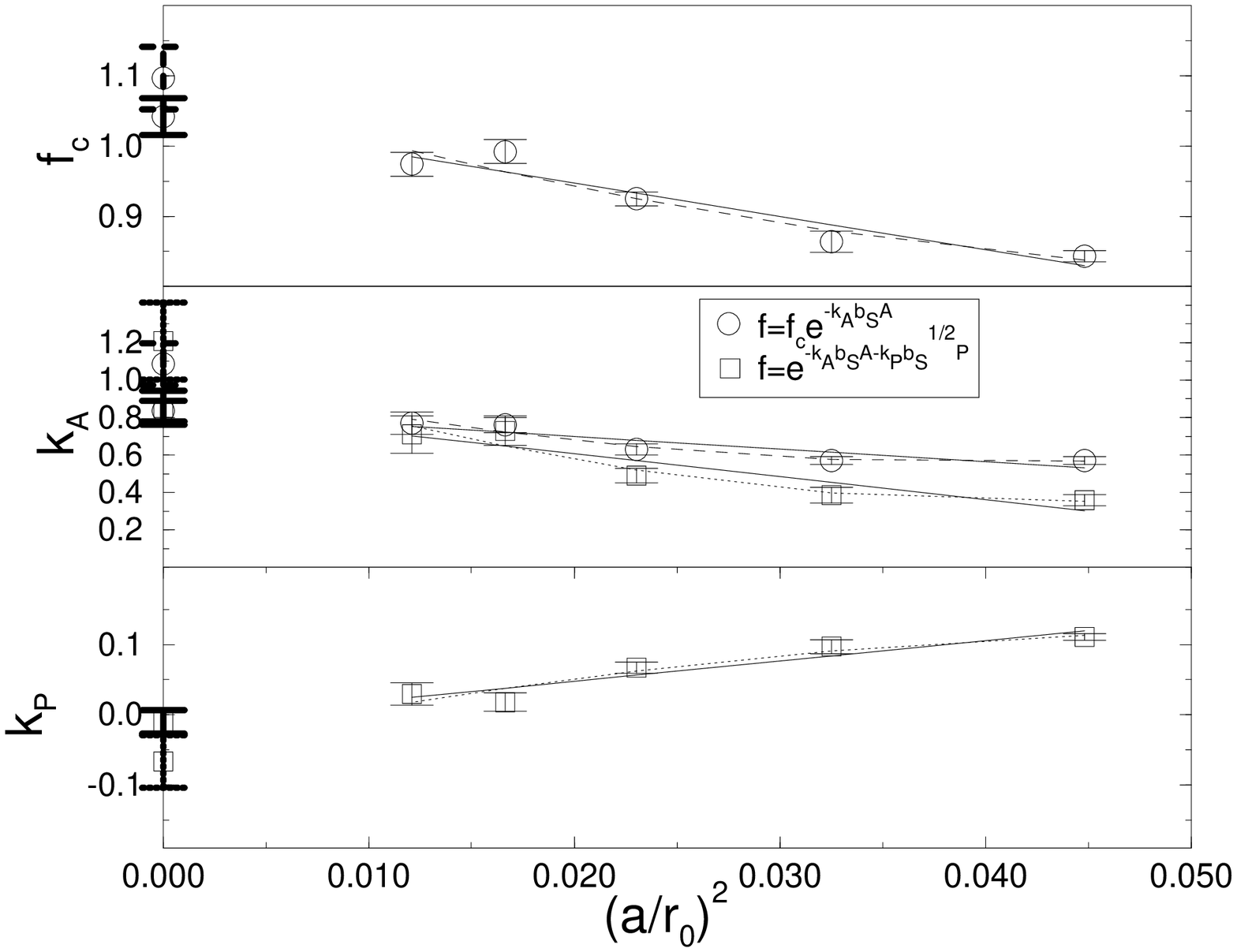}
\caption{$f_c$, $k_A$ and $k_P$ continuum extrapolations.}
\label{fsomfc}
\end{figure}

Typical values of $f$ are quite constant at different $\beta$'s, as shown in 
fig. \ref{fsomf}, having slopes from $-3.0(4)$ to $2.5(3)$. These slopes are a
reflection of the behaviour of the two-body potentials used to calculate the 
$f$'s. The parameters $f_c$, $k_A$ and $k_P$ are extrapolated in figure 
\ref{fsomfc}. 

The lattice artefacts of masses are known to behave roughly as $a^2$, but no 
theoretical justification exists for such behaviour of the artefacts of $f$. 
Therefore extrapolation results assuming this dependence should be given less
weight than fits of the model to continuum extrapolated energies presented 
below. These two approaches agree for values of $f$ in the continuum, while 
for the parameters of $f$ the quadratic extrapolations roughly agree 
with values obtained from continuum fits of the model as can be seen in 
table \ref{tlinearr}. 

\subsection{Continuum fit}

The fit data for all linear extrapolations is presented in table 
\ref{tlineard} and
for all quadratic extrapolations in table \ref{tquadd}.

Fit and parameter extrapolation results can be seen in table \ref{tlinearr}.
Fits were also performed with quadratic extrapolations for two-body potentials 
and linear for four-bodies and are denoted ``linear/quadratic'' in the table.
The parameter extrapolations have two $\chi^2$ values, one for each of the 
extrapolated parameters.

\begin{table}[htb]
\begin{center}
\begin{tabular}{|l|c|c|c|c|c|} \hline
$x/a,y/a,z/a$ & $f(E_0)$  & $E_0$[fm$^{-1}$] & $v_{s1}$[fm$^{-1}$] & 
$v_{s2}$[fm$^{-1}$] & $v_d$[fm$^{-1}$]  \\ \hline
2,2,0     & 0.92(6) & --0.515(25) & 7.44(34)  &           & 8.26(30)  \\
3,2,2     & 0.88(6) & --0.379(15) & 8.34(39)  & 8.432(40) & 9.14(40)  \\
3,3,0     & 0.77(2) & --0.439(10) & 8.38(39)  &           & 9.17(39)  \\
3,3,1     & 0.84(4) & --0.358(11) & 8.43(40)  & 8.55(40)  & 9.28(40)  \\
4,3,2     & 0.69(3) & --0.261(12) & 8.84(40)  & 9.07(40)  & 9.84(40)  \\
4,4,0     & 0.62(2) & --0.430(9)  & 8.98(31)  &           & 9.89(31) \\
4,4,1     & 0.65(2) & --0.376(15) & 9.08(40)  & 9.15(40)  & 10.02(40) \\
5,4,3     & 0.48(2) & --0.406(19) & 9.63(40)  & 9.63(40)  & 10.68(40) \\
5,5,0     & 0.46(2) & --0.388(14) & 9.55(30)  &           & 10.59(31) \\
5,5,1     & 0.48(2) & --0.354(20) & 9.63(40)  & 9.69(40)  & 10.71(40) \\
6,5,3     & 0.35(2) & --0.281(27) & 10.07(40) & 10.15(40) & 11.30(40) \\
6,6,0     & 0.33(2) & --0.334(19) & 10.06(31) &           & 11.25(32) \\
6,6,1     & 0.32(2) & --0.296(25) & 10.15(40) & 10.19(40) & 11.38(40) \\ \hline
\end{tabular}
\caption{Continuum fit data with linear extrapolations \label{tlineard}}
\end{center}
\end{table}

\begin{table}[htb]
\begin{center}
\begin{tabular}{|l|c|c|c|c|c|} \hline
$x/a,y/a,z/a$ & $f(E_0)$  & $E_0$[fm$^{-1}$] & $v_{s1}$[fm$^{-1}$] & 
$v_{s2}$[fm$^{-1}$] & $v_d$[fm$^{-1}$]  \\ \hline
2,2,0     & 0.90(16) & --0.40(11) & 9.29(17)  &           & 9.95(18) \\
3,2,2     & 0.86(8)  & --0.46(13) & 9.70(18)  & 9.73(18)  & 10.52(20) \\
3,3,0     & 0.75(11)  & --0.43(10) & 10.09(19) &           & 10.88(21) \\
3,3,1     & 0.68(5)  & --0.29(14) & 9.73(18)  & 9.89(19)  & 10.65(20) \\
4,3,2     & 0.55(6)  & --0.20(11) & 10.22(19) & 10.45(20) & 11.22(22) \\
4,4,0     & 0.56(9)  & --0.40(10) & 10.68(20) &           & 11.59(23) \\
4,4,1     & 0.52(5)  & --0.30(15) & 10.45(20) & 10.53(20) & 11.40(23) \\
5,4,3     & 0.37(4)  & --0.33(19) & 11.00(22) & 11.00(22) & 12.07(25) \\
5,5,0     & 0.35(7)  & --0.32(11) & 11.24(22) &           & 12.31(26) \\
5,5,1     & 0.36(4)  & --0.27(18) & 11.00(22) & 11.07(22) & 12.10(25) \\
6,5,3     & 0.27(4)  & --0.23(18) & 11.46(23) & 11.52(23) & 12.69(28) \\
6,6,0     & 0.19(5)  & --0.22(11) & 11.76(24) &           & 12.99(29) \\
6,6,1     & 0.21(4)  & --0.20(29) & 11.52(23) & 11.56(23) & 12.77(28) \\ \hline
\end{tabular}
\caption{Continuum fit data with quadratic extrapolations \label{tquadd}}
\end{center}
\end{table}

\begin{table}[htb]
\begin{center}
\begin{tabular}{|l|c|c|c|c|} \hline
type                      & $f_c$    & $k_A$      & $k_P$ & $\chi^2$ \\ \hline
all linear                & 1.08(3)  & 1.04(5)    & 0         & 0.8  \\ 
all linear, $f_c=1$       & 1        & 0.94(3)    & 0         & 1.3  \\
linear extrapol.          & 1.04(3)  & 0.83(6)    & 0         & 2.6/3.3 \\
linear extrapol. (II)     & 1        & 0.85(9)    & --0.02(2) & 2.3/2.6\\
linear/quadratic          & 1.04(4)  & 1.04(6)    & 0         & 0.7 \\
linear/quadratic, $f_c=1$ & 1        & 0.98(3)    & 0         & 0.8 \\
all quadratic             & 1.08(8)  & 1.39(13)   & 0         & 0.4  \\
all quadratic, $f_c=1$    & 1        & 1.27(6)    & 0         & 0.5  \\
quadratic extrapol.       & 1.10(5)  & 1.09(11)   & 0         & 2.7/0.4 \\ 
quadratic extrapol. (II)  & 1        & 1.2(2)     & --0.08(4) & 0.8/1.6 \\ 
\hline
\end{tabular}
\caption{Continuum fit and parameter extrapolation results. \label{tlinearr}}
\end{center}
\end{table}

All fits and extrapolations predict $f_c$ to be one or very slightly above.  
If it is set to one the fit to linear data and linear extrapolation give a 
$k_A$ somewhat below one, while the linear/quadratic fit has $k_A$ 
approximately one and the fit to quadratic data has a value slightly above 
unity. Since the linear/quadratic fits have low $\chi^2$ values and are 
most reliable for two-body potentials and four-quark binding energies, our 
best estimates for the continuum values are $f_c=1$, $k_A=1.0(1)$ and 
$k_P = 0$. The perimeter term should 
become unimportant in the continuum limit \cite{fur:95}, which is indeed 
observed. Thus the only parameter with physical significance is $k_A$. 

Continuum fits were also performed for squares extrapolated using the original
value $c=1.65$ in Sommer's equation \ref{esommer}. The resulting values of 
$f_c$ and $k_A$ for all linear and quadratic fits were within errors with fits
to extrapolations using $c=2.44$. As another test linear/quadratic and 
quadratic fits were performed with two-body potentials extrapolated with the 
$\beta=2.35$ point cut off, with similar results but higher $\chi^2$ values.

In the strong coupling limit
a factor $\exp{(-b_S A)}$ appears in the diagonal elements of the Wilson loop 
matrix, $A$ being the minimal surface bounded by straight
lines connecting the quarks. The factor is closely related to 
$f$ and originates from the sum of spatial plaquettes tiling the transition 
surface between two basis states. When moving to weaker couplings there exist 
correlated flux configurations for which the transition area is much smaller 
than $A$, which has been expected to lead to larger mixing among basis 
states \cite{mat:87}. If this is the case, it is not reflected in our best 
estimate of a $k_A$ about one. A larger transition area 
could be explained by the finite width of actual flux tubes instead of simply
lines as in strong coupling approximation. 

\section{How can the model be developed? \label{sdev}}

The failure of the simple $f$-model of eqs. \ref{emodel} and \ref{NV} 
to predict the energies of the tetrahedral
geometry has been proposed to be due to a dependence of the four-quark
energy on the first excited state(s) of the gluonic field in the static 
two-quark potential \cite{glpm:96}. A higher lying basis state would make 
the ground state attractive with increasing attraction when the size of the 
system gets larger, since the excited state gets closer to the ground state 
energy. Such a trend of increasing attraction is observed in the tetrahedral
energies unlike in the case of squares, whose energies get smaller with 
increasing size.

The lowest-lying gluonic excitation has non-zero angular 
momentum about the interquark axis, transforming in the $E_u$ representation
of the cubic symmetry group $D_{4h}$, while the second lowest excitation 
has the same $A_{1g}$ symmetry as the ground state. In our calculation the 
two-body paths used as a variational basis differ in the amount of fuzzing 
(a.k.a. smearing, blocking) applied to the lattice. Fuzzing only changes
the overlap with the excited states (to be denoted with a prime) possessing 
the $A_{1g}$ symmetry and does not bring in any of the lowest $E_u$ excited 
state. The question remains which one of these excited states has more effect 
on the tetrahedral energy.

The dependence of the 
four-quark energies on the $A_{1g}'$ excited states can be investigated using 
the fact that an 
increasing fuzzing level reduces the overlaps $C_i, i\ge 1$ of these excited 
states in 
the energy eigenstate expansion of the Wilson loop
\begin{equation}
<W(R,T)> = \sum_n C_n(R)e^{-V_n(R)T}.
\end{equation}
The overlaps $C_n(R) \ge 0$ have a normalization condition
$$
\sum_n C_n(R) = 1.
$$
Increasing the fuzzing level used in the calculation of four-body energies
should worsen the convergence of any four-body energy with a significant 
dependence on the excited
state(s) with the symmetry of the ground state. Our group has been using 
fuzzing level 20 with the factor multiplying the link to be fuzzed $c=4$, 
while a group 
in Wuppertal doing related work has chosen 150 with $c=2$ to maximize the 
ground state overlap. At $\beta=2.5$ they obtain $C_0$ values from 0.81 to 
0.98 on a $16^4$ lattice. 

We made runs at $\beta=2.5$ on a $24^3\times32$ lattice with three different 
fuzzing levels used to calculate the four-quark energy of a regular 
tetrahedron. 
Ground state overlaps $C_0$ of two-quark paths at 
different separations $r/a,d/a$ and at different fuzzing levels are shown
in table \ref{tfuz}. The four-body operators are 
constructed from the same paths, and the relative quality of convergence of 
the binding energy $E_4$ as a series in $T$ is also shown.

\begin{figure}[htb]
\hspace{0cm}\epsfxsize=400pt\epsfbox{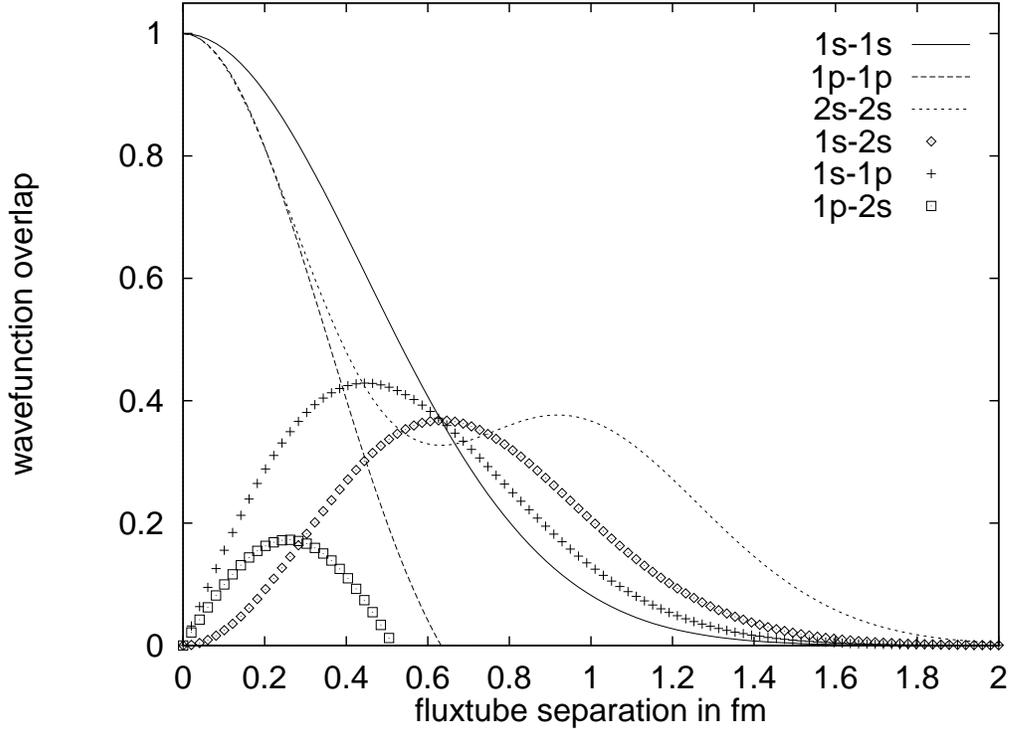}
\caption{Overlaps of fluxtube wavefunctions as functions of fluxtube 
separation in the Isgur-Paton model with N=2. \label{fol}}
\end{figure}

The table shows that best convergences of the four-quark energies are 
obtained when the two-body paths have practically no overlap with gluonic 
excitations. Therefore the only way excitations can contribute is through
overlaps of two-quark paths with excited states of gluon fields between
{\em other} quark pairs. This contribution is likely to be significant; e.g. 
the Isgur-Paton string model \cite{isg:85} with N=2 predicts overlaps 
between 1s, 1p (corresponding to $A_{1g}$, $E_u$) and 1s, 2s (corresponding to 
$A_{1g}$, $A_{1g}'$) states to be $\approx 0.4$ when the centres of two 
parallel 
fluxtubes are separated by $\approx 0.4$ fm and $\approx 0.7$ fm respectively.
This can be seen from fig. \ref{fol}. The rapid worsening of convergence in 
table \ref{tfuz} with decreasing ground-state overlap and increasing overlap 
with $A_{1g}'$ excitations suggests that the lowest-lying gluonic 
excitation with $E_u$ symmetry has a more important effect on the four-quark
energies than the higher lying $A_{1g}'$ excitations.

\begin{table}[htb]
\begin{center}
\begin{tabular}{|l|c|c|c|} \hline
          & FL=150     & FL=20   & FL=0  \\ \hline
$r/a,d/a$ & $C_0$ \%   & $C_0$ \%   & $C_0$ \%  \\ \hline
2,2       & N/A        & 100.0(1)   & 98.2(1)  \\
          & interm.    & best       & worst \\ \hline
3,3       & 98.7(1)   & 100.0(1)    & 96.7(1)   \\
          & interm.   & best        & worst \\ \hline
4,4       & 98.2(2)   & 99.7(1)    & 94.3(1)  \\
          & ``best''      & ``best''       & worst \\ \hline
5,5       & 97.9(2)   & 99.5(1)    & 91.1(1)   \\
          & interm.   & best       & worst \\ \hline
6,6       & 98.0(2)   & 99.1(1)    & 85.3(1)   \\
          & interm.   & best       & worst \\ \hline
7,7       & 99.1(4)   & 98.0(1)    &  N/A  \\ 
          & interm.   & best       & worst \\ \hline
\end{tabular}
\caption{Ground state overlaps and $E_4$ convergence at different fuzzing 
levels (FL) and interquark distances. \label {tfuz}}
\end{center}
\end{table}

The simple $f$-model of eqs. \ref{emodel} and \ref{NV} can now be extended
by using (in addition to the $A,B,C$ shown in figure \ref{fconf}) three 
additional basis states describing the same quark partitions but involving 
the first excited state 
of the two-body potential. The resulting $6 \times 6$ matrices contain
two new gluon field overlap factors $f^a$ and $f^c$ similar to $f$, which  
however measure the overlap between an excited and ground state basis state 
and two excited basis states respectively. Solving the determinant analogous 
to eq. \ref{emodel} gives for a regular tetrahedron two degenerate and two 
non-degenerate energy eigenvalues. The degenerate negative ground state energy
of a regular tetrahedron would correspond to the lower of the degenerate 
eigenvalues, while it is harder for the model to predict the third 
eigenenergy from simulations since it is dominated by gluonic excitations as 
argued in ref. \cite{glpm:96}. 

To estimate these eigenvalues, a parameterization for $f$ from table 
\ref{t2parfc} can be used with the minimal transition areas for regular 
tetrahedra calculated in ref. \cite{fur:96}. A parameterization for the 
energy of the first excited two-body state is found from ref. \cite{gre:96}. 
Because all the quarks in a regular tetrahedron 
have the same distance from each other, all the excited basis states have the 
same energy. Using these values, the simulation results for regular tetrahedra
and a self consistency argument allows estimation of other parameters 
of the model. 

The values obtained give the lower degenerate eigenvalue as the 
lowest eigenvalue when $f^c > 1$ because of a non-degenerate eigenvalue being 
unstable around $f_c = 1$. 
This unstable eigenvalue is probably stabilized by a minor contribution from 
the $A_{1g}'$ excited state, requiring treatment with 9 basis states.

The values for the new overlap 
factor $f^a$ have a similar linear behaviour with the size of the system as is 
observed in figure \ref{fol} in the overlap of 1s-1p Isgur-Paton fluxtubes 
with separation in the same distance range, supporting reliability of the 
estimation scheme. 

Work is in progress on applying this extended model to geometries more 
complicated than the regular tetrahedron \cite{pp:96b}.

\section{Conclusions}

The conclusions in this work can be summarized as follows:
\begin{enumerate}
\item{Choosing $c=2.44$ (corresponding to $r_0 \approx 0.66$ fm) in eq. 
\ref{esommer} leads to better agreement of various 
continuum potential models with each other, and in our case gives lattice 
spacings that agree with those obtained from the string tension. This value of
$c$ thus seems a better choice than the original choice $c=1.65$ ($r_0 \approx 
0.49$ fm) by Sommer, although it cannot be excluded that the difference is 
accidental due to uncertainties in the potential models and the quenched SU(2) 
string tension.}
\item{The lattice spacings $a$ at the six values of $\beta$ used in 
simulations give the $\beta$-function $b\equiv \frac{\partial \beta}{\partial 
\ln a}$ in agreement with ref. \cite{eng:95} when $\beta\geq 2.35$. At 
$\beta=2.5$ the measured value accounts for 80\% of the three-loop 
perturbative prediction, presenting a very modest improvement of 2\% from the 
two-loop case.}
\item{A quadratic extrapolation is preferred for two-body potentials, whose 
continuum parameterization is given in section \ref{2exp}. Four-quark binding 
energies are scaling from $\beta \ge 2.35$ and their values do not change 
significantly in the range of $\beta$ values simulated. In practice the 
simulation results for the binding energies can thus be used directly, whereas
it is recommended to use the continuum parameterization for two-body potentials
to avoid introducing lattice artefacts into a four-quark energy model.}
\item{Parameter extrapolations and continuum fits of the simple $f$-model show
that in the continuum, the normalization of the gluon field overlap factor 
$f$ can be safely set to one and the perimeter term to zero, leaving 
the constant $k_A$ multiplying the area as the only physically relevant 
parameter. Its value is about one and not $\ll 1$ as predicted 
for the transition area in ref. \cite{mat:87}.}
\item{In our simulations, the effect on the four-quark energies of excited 
states of the gluon field between two quarks comes from the overlap of 
two-body paths with excited two-body fields of other quark pairings. An 
extended $f$-model using the excited states of the two-body potentials 
can reproduce the negative, 
degenerate ground state energy of a regular tetrahedron unlike the simple 
model based on ground state potentials.}
\end{enumerate}

Other possible extensions of the simple $f$-model include instanton 
effects (ref. \cite{sch:96}) and four-body interactions in the strong 
coupling limit (ref. \cite{gro:88}).

A study using microscopical flux distributions would be helpful in determining
the relation of the shape of two- and four-quark fluxtubes and their overlaps 
to the parameterization of $f$. Our next project attempts this by using 
sum-rules similar to those derived by C. Michael \cite{mic:95}.

\section{Acknowledgement}

I warmly thank A.M. Green for his support and encouragement. 
I would like to thank J. Lukkarinen and P. Laurikainen for discussions, 
C. Schlichter from Wuppertal for kindly giving us their simulation results and 
S. Furui and B. Masud for communicating us their results of transition 
areas for tetrahedra. Funding from the Finnish Academy is gratefully 
acknowledged. Our simulations were performed on the Cray C94 at the Center 
for Scientific Computing in Espoo.

\end{document}